\documentclass{PoS}
\usepackage{amsmath}

\title{Low energy spectra in many flavor QCD 
with $N_f=$12 and 16}
\ShortTitle{Low energy spectra in many flavor QCD with $N_f=$12 and 16}
\author{
Yasumichi Aoki$^a$,
Tatsumi Aoyama$^a$, 
Masafumi Kurachi$^a$,
Toshihide Maskawa$^a$, 
Kei-ichi Nagai$^a$, 
\speaker{Hiroshi Ohki}$^a$, 
Akihiro Shibata$^b$,  
Koichi Yamawaki$^a$ and 

Takeshi Yamazaki$^a$

\hspace*{55mm} LatKMI Collaboration
\\ \\
$^a$
Kobayashi-Maskawa Institute for the Origin
of Particles and the Universe (KMI), Nagoya University, Nagoya
464-8602, Japan\\
$^b$
Computing Research Center, High Energy Accelerator Research Organization (KEK), 
Tsukuba 305-0801, Japan

E-mail: \email{ohki@kmi.nagoya-u.ac.jp}}

\abstract{
We present our result of the many-flavor QCD. 
Information of the phase structure of many-flavor SU(3) gauge theory is of great interest, 
since the gauge theories with the walking behavior near the infrared fixed point are candidates 
of new physics for the origin of the dynamical electroweak symmetry breaking. 
We study the SU(3) gauge theories with 12 and 16 fundamental fermions. 
Utilizing the HISQ type action which is useful to study the continuum physics,
we analyze the lattice data of the mass and the decay constant of 
the pseudoscalar meson and the mass of the vector meson as well  
at several values of lattice spacing and fermion mass.
The finite size scaling test in the conformal hypothesis is also performed.
Our data is consistent with the conformal scenario for $N_f=12$.
We obtain the mass anomalous dimension 
$\gamma_m \sim 0.4-0.5$. 
An update of $N_f=16$ study is also shown.
}

\FullConference{The 30 International Symposium on Lattice Field Theory - Lattice 2012,\\
		June 24-29, 2012\\
		Cairns, Australia}

\begin{document}

\section{Introduction}
There has been a renewed interest in 
the study of 
QCD with large number of the massless fermions in the fundamental representation (``large $N_f$ QCD'') 
in the context of walking technicolor having approximate scale invariance in the infrared (IR) region
and large anomalous dimension $\gamma_m \simeq 1~$\cite{Yamawaki:1985zg}.
Such an approximate scale-invariant dynamics may in fact be realized in the large $N_f$ QCD: 
The perturbative two-loop beta function predicts a 
non-trivial infrared fixed point $\alpha_*$ ($0<\alpha_*<\infty$) in the range of $9 \le N_f \le 16$ 
in the asymptotically free SU(3) gauge theory
~\cite{Caswell:1974gg, Banks:1981nn}.
As a powerful tool of a nonperturbative study,  one uses  
lattice QCD simulations, which  can 
in principle determine the phase structure 
of the SU(3) gauge  
theories with various number of fermions.
In addition to the pioneering works~\cite{Iwasaki:1991mr, Brown:1992fz, Damgaard:1997ut}, 
there are many lattice works in the large $N_f$ QCD in the recent years.
In particular, the system of the $N_f=12$ has been widely investigated 
by the lattice approach,
such as running coupling, lattice phase diagram, low-energy spectra 
and so on. (See, for a review, Ref.~\cite{plenary}.)

We investigate the 12 and 16-flavor SU(3) gauge theories using a variant of 
the highly improved staggered quark (HISQ) action~\cite{Follana:2006rc}
to reduce the discretization error.
We study several bound-state masses such as the pseudoscalar meson $\pi$ and vector meson $\rho$ as well as the decay 
constant of $\pi$, by varying the fermion bare mass $m_f$.
In this work, we introduce a quantity for the scaling test of the conformal hypothesis with finite volume in the analyses 
for $N_f=12$. 
Using this quantity, we can analyze the data without any assumption of the fitting form 
in the scaling test.
We also discuss possible finite size and mass corrections in the scaling.
We find that our results of $N_f=12$ are consistent with hyperscaling with $\gamma=0.4-0.5$.
These results of $N_f=12$ have already been published in Ref.~\cite{Aoki:2012eq}.

Besides the $N_f=12$ theory, we also simulate the $N_f=16$ theory with the same setup. 
Since this theory is expected to deeply reside in the conformal phase,
it is helpful to understand conformal signals from numerical simulations.
The preliminary results of this theory are also shown.

\section{$N_f=12$}

We use a version of the HISQ~\cite{Follana:2006rc} action for many-flavor simulations 
but without the tadpole improvement and the mass correction term for heavy fermions.
Gauge configurations are generated by HMC algorithm using MILC code ver.7 
with various parameter sets for the fermion mass $m_f$, 
volume and the bare coupling $\beta = 6/g^2$.   
We calculate several bound-state masses, such as the pseudoscalar meson $\pi$
and vector meson $\rho$, and the decay constant of $\pi$.
If the theory is in the conformal window, the hadron mass $M_H$ and $\pi$ 
decay constant $F_\pi$ obey the conformal hyperscaling 
\begin{equation}
 M_H \propto m_f^{\frac{1}{\gamma_*+1}},\; 
  F_\pi \propto m_f^{\frac{1}{\gamma_*+1}},
  \label{eq:hyperscaling}
\end{equation}
where $\gamma_*$ denotes the mass anomalous dimension at the
IR fixed point.
On the other hand, if the theory is in the phase of chiral symmetry breaking,  
the leading fermion mass dependence of $\pi$ is
\begin{equation}
 M_\pi^2 \propto m_f,\; 
  F_\pi = c_0 + c_1 m_f,
  \label{eq:chpt_leading}
\end{equation}
with $c_0\ne 0$,
and the vector meson mass does not vanish in the chiral limit.

The spectra obtained in our lattice simulation will be tested against 
these two hypotheses in this work.

\subsection{Primary analysis}
\begin{figure}[htbp]
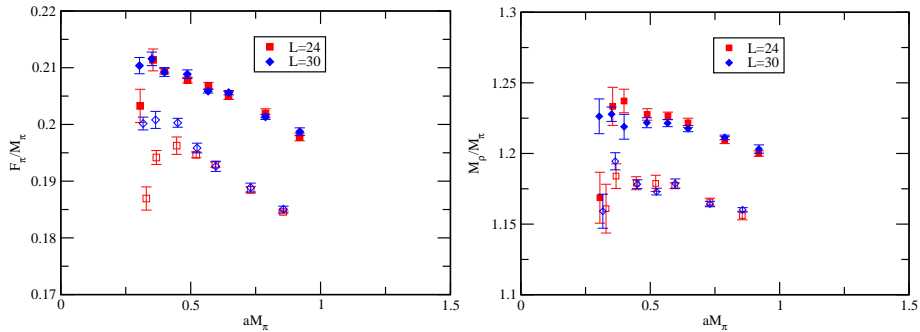

\centering
   \includegraphics[width=6.cm,clip]{fpi-pi_ratio_r.eps}
   \includegraphics[width=6.cm,clip]{rho-pi_ratio_r.eps} 
 \caption{Dimension-less ratios $F_\pi/M_\pi$ and $M_\rho/M_\pi$
 as functions of $aM_\pi$ for $N_f=12$ at $\beta=3.7$ (filled symbol) and
 $4.0$ (open symbol) for two largest volumes.}
 \label{fig:ratio_mpi_nf12}
\end{figure}

For a primary analysis dimensionless ratios, $F_\pi/M_\pi$ and $M_\rho/M_\pi$,
are plotted against $aM_\pi$ in Fig.~\ref{fig:ratio_mpi_nf12}.
The left panel plots $F_\pi/M_\pi$
on the largest two volumes at the two bare gauge couplings $\beta=3.7$ and $4$.
If we look at the results for $\beta=3.7$ (filled symbols),
$F_\pi/M_\pi$ tends to be flat for smaller $\pi$ masses,
which shows clear contrast to ordinary QCD case.
The behavior is consistent with the hyperscaling in Eq.~(\ref{eq:hyperscaling}). 
The $M_\pi$ dependence of the ratio
at the larger mass can be realized by the correction to the hyperscaling
which may be different from one quantity to another.
For $\beta=4$ (open symbols) there is no flat range without volume dependence.
Similar observation can be made for the other ratio $M_\rho/M_\pi$
shown in the right panel of Fig.~\ref{fig:ratio_mpi_nf12}.
The flattening is observed for $\beta=3.7$ again, but
the range is wider than $F_\pi/M_\pi$. In this case $\beta=4$ shows 
the flattening, too.
The difference of the constant in the flat region could be caused by a discretization
effect.

Existence of the scaling for $F_\pi$ at $\beta=3.7$ and absence
at $\beta=4$ at the same $aM_\pi$ can be made possible 
if the $M_\pi$ in the 
physical unit is larger (thus the correction is no longer negligible)
for $\beta=4$, {\it i.e.}, the lattice spacing decreases as $\beta$
increases. In that case the physical volume is smaller for $\beta=4$,
which gives a reason for the volume effect observed only for $\beta=4$.
Actually a crude analysis of two lattice spacing matching shows the 
result  $a(\beta=3.7) > a(\beta=4)$ which is consistent 
with being in the asymptotically free domain.

\subsection{Finite-size hyperscaling}

In the conformal window with finite masses and volume, 
the renormalization group analysis tells us that the scaling behavior 
for low-energy spectra 
which should obey the universal scaling relations~\footnote{
For reviews,  see e.g. 
\cite{DeGrand:2009mt,DelDebbio:2010ze}.
} as
\begin{equation}
  \xi_p \equiv L M_{p} = f_p(x), \quad   \xi_F \equiv L F_{\pi} =f_F(x),
 \label{eq:fss_mass}
\end{equation}
where the subscript $p$ distinguishes the bound state, $p=\pi$ or $\rho$ in this study.
The product of bound state mass or decay constant and linear system
size falls into a function of a single scaling variable  
$x=L\cdot m_f^{\frac{1}{1+\gamma_*}}$, 
where $\gamma_*$ is the mass anomalous dimension at the IR
fixed point. 
We shall call the scaling relation the finite-size 
hyperscaling (FSHS).
While the forms of the scaling functions $f_p(x)$ are unknown in general, 
the asymptotic form should be $f(x) \sim x$ at large $x$
because it must reproduce the hyperscaling relation Eq.~(\ref{eq:hyperscaling})
in large volumes.

\begin{figure}[tb]
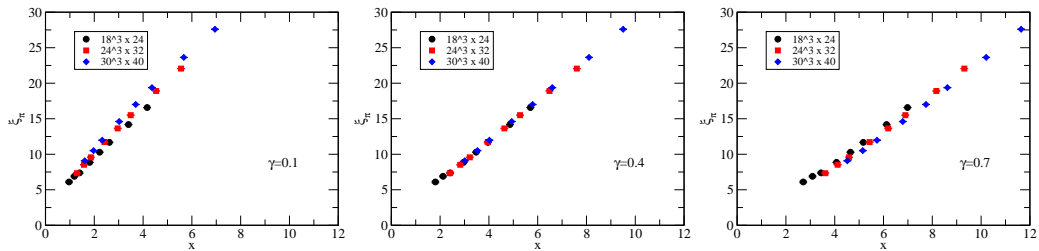

\centering
      \includegraphics[width=4.5cm,clip]{b3.7_mpi_g=0.1.eps}
      \includegraphics[width=4.5cm,clip]{b3.7_mpi_g=0.4.eps}
      \includegraphics[width=4.5cm,clip]{b3.7_mpi_g=0.7.eps}
 \caption{$\xi_\pi$ is plotted as a function of the scaling variable $x$ for 
 $\gamma=0.1$, $0.4$, and $0.7$ from left to right for $N_f=12$ at
 $\beta=3.7$.  An alignment is seen for $\gamma\sim 0.4$.
  }
  \label{fig:mpi_g}
\end{figure}

Now we examine whether our data for the bound state masses and decay constant 
obey FSHS. 
To visualize how the scaling works 
we show $\xi_\pi$ as functions of $x = L \cdot m_f^{1/(1+\gamma)}$ 
for several values of $\gamma$ in Fig.~\ref{fig:mpi_g}.
It is observed that the data align well
with around $\gamma=0.4$, while they become scattered for $\gamma$ 
away from that value. This indicates the existence of possible FSHS with 
$\gamma\sim 0.4$.
A similar alignment is observed for $\xi_F$ as well.
In this case one finds the optimal scaling at around $\gamma\sim 0.5$.

To quantify the ``alignment'' we introduce an evaluation function 
$P(\gamma)$ for an observable $p$ as follows. 
Suppose $\xi^j$ be a data point of the measured observable $p$ at 
$x_j = L_j \cdot m_j^{1/(1+\gamma)}$ and $\delta\xi^j$ be the error of $\xi^j$. 
$j$ labels distinction of parameters $L$ and $m_f$. 
Let $K$ be a subset of data points $\{(x_k, \xi_k)\}$ from which 
we construct a function $f^{(K)}(x)$ that represents the subset of data. 
Then the evaluation function is defined as 
\begin{equation}
  \label{eq:p}
  P(\gamma) =
  \frac{1}{\mathcal{N}}
  \sum_{L}
  \sum_{j \not\in{K_L}}
  \frac{\left|\xi^{j} - f^{(K_L)}(x_j)\right|^2}
       {\left|\delta \xi^{j}\right|^2}, 
\end{equation}
where $L$ runs through the lattice sizes we have,
the sum over $j$ is taken for a set of data points that do not 
belong to $K_L$ which includes all the data obtained on the lattice 
with size $L$.
$\mathcal{N}$ denotes the total number of summation. 
Here we choose for the function $f^{(K_L)}$ a linear interpolation 
of the data points of the fixed lattice size $L$ for simplicity, 
which should be a good approximation of $\xi$ for large $x$. 
In the evaluation function Eq.~(\ref{eq:p}), the data points need to be 
taken for a range of $x = L \cdot m_f^{1/(1+\gamma)}$ 
in which there is an overlap of available data 
for all volumes, $L=18$, $24$, and $30$ 
within the range $[x_\text{min}, x_\text{max}]$.
We take the value of $x_\text{min}$ ($x_\text{max}$) 
as the smallest (largest) $m_f$ for the largest (smallest) 
volume $L$ in our simulation parameters.
Note, however, we may need to incorporate some neighboring data 
outside this range to obtain the interpolated value $f^{(K)}(x)$ 
by the spline functions.

\begin{figure}[tb]
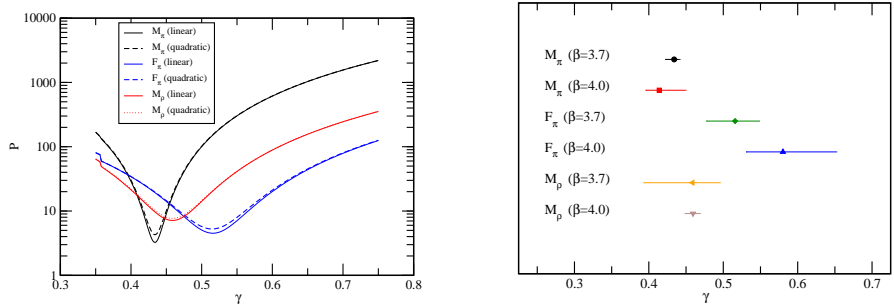

\centering
 \includegraphics[width=5.5cm,clip]{b3.7_P_io_dep.eps} \quad \quad \quad
 \includegraphics[width=5cm,clip]{gamma_asym.eps}
\caption{
(Left) The $\gamma$ dependence of the evaluation function $P$ 
for $M_\pi$, $F_\pi$, and $M_\rho$ at $\beta=3.7$ is plotted. 
The vertical axis shows the values of $P$ at each of $\gamma$ where 
the three volumes and full range of $x$ for the data are considered. 
The solid and dashed curves show the results of $P(\gamma)$ 
with the interpolation functions $f(x)$ 
by the linear and quadratic functions, respectively. 
(Right)
The results of the values of $\gamma$ for three observables at 
two $\beta$ are summarized, 
where the statistical and systematic errors are added in quadrature. 
}
\label{fig:P_io}
\end{figure}

The evaluation function for all the quantities, $M_\pi$, $M_\rho$ and 
$F_\pi$, is plotted in the left panel of Fig.~\ref{fig:P_io}. 
A clear minimum is observed at which the optimal alignment of the data 
is achieved. 
It is noted that the value of $P(\gamma)$ is $\mathcal{O}(1)$ at the minimum. 
The systematic error due to the ambiguity of the interpolation is 
estimated by the difference of the optimal $\gamma$'s obtained 
with linear and quadratic spline interpolations. 
The comparison of these $P(\gamma)$'s is also shown in the left panel of Fig.~\ref{fig:P_io}. 
The minima for the quadratic spline interpolation appear approximately 
at the same place as those for the linear one. 
It is found that this systematic error is always smaller than the statistical error.
The other uncertainties due to the finite size and mass effects are estimated 
by the variations of optimal $\gamma$ with respect to the change of 
both the x-range and $L$ used in the analyses.
The results with all the errors added in quadrature are summarised 
in the right panel of Fig.~\ref{fig:P_io}.  
The details of the analysis are shown in Ref.~\cite{Aoki:2012eq}.
All the results are consistent with each other within $2 \sigma$ level, 
except for $\gamma$ from $F_\pi$ at $\beta=4$ 
for which the scaling region was suspected to be outside of the 
parameter range we have examined in the previous subsection.
From these analyses, we conclude that our data for the $N_f=12$ 
theory are reasonably consistent with the FSHS.
The resulting mass anomalous dimensions is 
$0.4 \le \gamma_* \le 0.5$.

In the test of the chiral broken scenario, 
it turns out that  the natural chiral expansion parameter 
$\chi=N_f\left( \frac{M_\pi}{ 4\pi F_\pi/\sqrt{2} }\right)^2$ is 
very large in the region we simulated, 
which is evaluated as $\chi \sim 39$ at the smallest $M_\pi$ using the value of $F_\pi$ in the chiral limit.
With this large $\chi$, we could not consistently analyze the data based on the ChPT.
Further efforts would be required to arrive at a decisive conclusion.

\section{$N_f=16$}

In the previous report~\cite{Aoki:2012ep} we presented that the result of the $\xi_\pi$ is consistent with
the FSHS, but the optimal $\gamma$ decreases as $\beta$ increases in $\beta \le 3.5$.
Furthermore the results of the $\gamma$ is much larger than the perturbative result,
$\gamma \sim 0.025$.
In this theory the perturbation would be reliable because of the small coupling constant at the 
IR fixed point, so that we considered that our result contains large systematic errors.
Thus, we investigate the $\beta$ dependence of $\gamma$ in larger $\beta$ region
than the ones we simulated in the previous report.

Using the same simulation setup as in $N_f=12$, 
we perform simulations at several values of $\beta$ adding to the previous work,
such as 5 and 12, on various spatial volumes, $L=8, 12, 16, 24$ and 30.
The range of the fermion mass is $0.03 \le m_f \le 0.2$, and
the typical length of the trajectory is roughly 1000. 

\begin{figure}[!t]
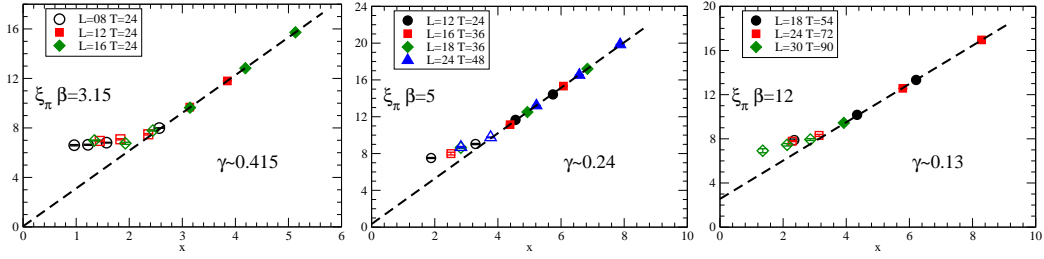

\centering
\includegraphics*[angle=0,width=0.3\textwidth]{mpi_gfit_b03.15.eps}
\includegraphics*[angle=0,width=0.3\textwidth]{mpi_gfit_b05.00.eps}
\includegraphics*[angle=0,width=0.3\textwidth]{mpi_gfit_b12.00.eps}
\caption{
$\xi_\pi$ in the 16 flavors at $\beta = 3.15$(left), 5(center) and $12$(right).
The different symbols denote the data at the different volumes.
The dashed line denotes the fit result of the finite-size hyperscaling.
The filled and open symbols represent the data included in the fit 
and omitted from the fit, respectively.
\label{fig:4}
}
\end{figure}

\begin{figure}[!t]
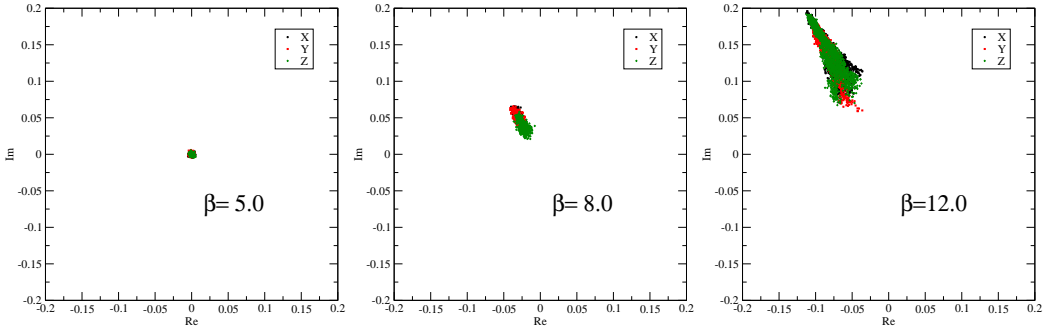

\centering
\includegraphics*[angle=0,width=0.3\textwidth]{b05_PL.eps}\
\includegraphics*[angle=0,width=0.3\textwidth]{b08_PL.eps}\
\includegraphics*[angle=0,width=0.3\textwidth]{b12_PL.eps}\
\caption{
Scatter plot of the Polyakov loops in the spacial directions.
}\label{fig:5}
\end{figure}

The plots in Fig.~\ref{fig:4} show that for $\beta = 5$ and 12
the $\xi_\pi$ aligns well as a function of $x$ using an optimal value of the $\gamma$
as in the $N_f = 12$ case.
The value, however, largely depends on $\beta$.
At the highest $\beta$ the $\gamma$ is still five times larger than the perturbative result,
although the result at this $\beta$ would include large systematic error coming from 
finite volume.
Fig.~\ref{fig:5} shows the scatter plots of Polyakov loop in the spatial directions 
with fixed bare mass $m_f=0.3$ and lattice sizes $L=18, T=32$.
As shown here, at the highest $\beta$, the Polyakov loop has a non-zero value 
associated with the center symmetry breaking. On the other hand, 
at the lower $\beta$ region, this value decreases with $\beta$.
We consider that this symmetry breaking at the higher $\beta$ occurs due to 
the small physical volume, so that to reduce finite volume effects we would need
much larger lattice size than $L=18$ at $\beta = 12$.
We will continue investigations with larger volumes at higher $\beta$ region
to obtain the correct value of the $\gamma$ at the vicinity of the infrared fixed point.


\section{Summary and outlook}
We have studied the SU(3) gauge theories with the fundamental 
12 and 16 fermions using a HISQ type staggered fermion action, 
For the 12-flavor case, 
we attempt to determine the phase of this theory through the analysis of 
$M_\pi$, $M_\rho$ and $F_\pi$. 
Our present data is consistent with the conformal hypothesis.
The mass anomalous dimension, $\gamma_*$ at the infrared fixed point was estimated 
through the (finite-size) hyperscaling analysis. 
Our result is $\gamma_* \sim 0.4 - 0.5$, which is not as big as $\gamma_* \sim 1$ 
for the theory to be close to the realistic technicolor model. 
In the test of chiral broken scenario, the chiral expansion parameter $\chi$ is 
much larger than one even at the smallest $M_\pi$.
We could not consistently analyze the data based on the chiral expansion. 
A possibility of the chiral broken phase in $N_f=12$ is not excluded yet.
More detailed analyses with more data at larger volume and lighter mass would be required.
For the 16-flavor case, 
the pion mass data exhibit the scaling which is consistent with the conformal scenario,  
while the obtained value of the $\gamma$ is much bigger than the perturbative result. 
To obtain the $\gamma$ at the infrared fixed point, further study of the $\gamma$,
especially volume dependence would be required.
Another series of the study in our project has been reported for the test of the walking behavior in $N_f=8$~\cite{Nf=8}, 
and an analytical calculation of the hyperscaling through the Schwinger-Dyson equation~\cite{Aoki:2012ve}.
\section*{Acknowledgments}
Numerical calculations have been carried out
on the cluster system $\varphi$ at KMI, Nagoya University.
This work is supported in part by JSPS Grants-in-Aid for Scientific
Research (S) No.~22224003, (C) No.~21540289 (Y.A.), (C) No.~23540300 (K.Y.),
(C) No.~24540252 (A.S.)
and 
Grants-in-Aid of the Japanese Ministry for Scientific Research on
Innovative 
Areas No.~23105708 (T.Y.).


\begin{thebibliography}{99}

\bibitem{Yamawaki:1985zg}
  K.~Yamawaki, M.~Bando and K.~-i.~Matumoto,
  Phys.\ Rev.\ Lett.\  {\bf 56}, 1335 (1986).

  
\bibitem{Caswell:1974gg}
  W.~E.~Caswell,
  Phys.\ Rev.\ Lett.\  {\bf 33} (1974) 244.
  
\bibitem{Banks:1981nn}
  T.~Banks and A.~Zaks,
  Nucl.\ Phys.\ B {\bf 196} (1982) 189.
  
\bibitem{Iwasaki:1991mr}
  Y.~Iwasaki, K.~Kanaya, S.~Sakai and T.~Yoshie,
  Phys.\ Rev.\ Lett.\  {\bf 69} (1992) 21.

\bibitem{Brown:1992fz}
  F.~R.~Brown, H.~Chen, N.~H.~Christ, Z.~Dong, R.~D.~Mawhinney, W.~Schaffer and A.~Vaccarino,
  Phys.\ Rev.\ D {\bf 46} (1992) 5655
  [hep-lat/9206001].
  
\bibitem{Damgaard:1997ut}
  P.~H.~Damgaard, U.~M.~Heller, A.~Krasnitz and P.~Olesen,
  Phys.\ Lett.\ B {\bf 400} (1997) 169
  [hep-lat/9701008].
  

\bibitem{plenary}
J.~ Giedt, plenary talk at Lattice 2012, PoS LATTICE {\bf 2012} (2012) 006;
  E.~T.~Neil,
  PoS LATTICE {\bf 2011}, 009 (2011)
  [arXiv:1205.4706 [hep-lat]], and references therein.

\bibitem{Follana:2006rc}
  E.~Follana {\it et al.}  [HPQCD and UKQCD Collaborations],
  Phys.\ Rev.\ D {\bf 75} (2007) 054502
  [hep-lat/0610092].
  
  
\bibitem{Aoki:2012eq}
  Y.~Aoki, T.~Aoyama, M.~Kurachi, T.~Maskawa, K.~-i.~Nagai, H.~Ohki, A.~Shibata, K.~Yamawaki,
  T.~ Yamazaki, (LatKMI collaboration),
  Phys.\ Rev.\ D {\bf 86} (2012) 054506
  [arXiv:1207.3060 [hep-lat]].

  
\bibitem{DeGrand:2009mt}
  T.~DeGrand and A.~Hasenfratz,
  Phys.\ Rev.\ D {\bf 80} (2009) 034506
  [arXiv:0906.1976 [hep-lat]].

\bibitem{DelDebbio:2010ze}
  L.~Del Debbio and R.~Zwicky,
  Phys.\ Rev.\ D {\bf 82} (2010) 014502
  [arXiv:1005.2371 [hep-ph]].

\bibitem{Aoki:2012ep} 
  Y.~Aoki, T.~Aoyama, M.~Kurachi, T.~Maskawa, K.~-i.~Nagai, H.~Ohki, A.~Shibata,
K.~Yamawaki, T.~Yamazaki, (LatKMI collaboration), 
  PoS LATTICE {\bf 2011}, 080 (2011)
  [arXiv:1202.4712 [hep-lat]].

\bibitem{Nf=8}
  Y.~Aoki, T.~Aoyama, M.~Kurachi, T.~Maskawa, K.~-i.~Nagai, H.~Ohki, A.~Shibata, K.~Yamawaki,
  T.~Yamazaki, (LatKMI collaboration),
talk at Lattice 2012, PoS LATTICE {\bf 2012} (2012) 035.

\bibitem{Aoki:2012ve}
  Y.~Aoki, T.~Aoyama, M.~Kurachi, T.~Maskawa, K.~-i.~Nagai, H.~Ohki, A.~Shibata, K.~Yamawaki,
  T.~Yamazaki, (LatKMI collaboration), 
  Phys.\ Rev.\ D {\bf 85} (2012) 074502
  [arXiv:1201.4157 [hep-lat]]; talk at Lattice 2012, PoS LATTICE {\bf 2012} (2012) 059.
  
\end{thebibliography}
\end{document}